\documentclass[manuscript, screen]{acmart}
\AtBeginDocument{%
  }

\setcopyright{none}

\acmConference{}

\setcopyright{none}

\usepackage[justification=centering]{caption}

\begin{document}

\title{Inject, Fork, Compare: Defining an Interaction Vocabulary for Multi-Agent Simulation Platforms}

\author{HwiJoon Lee}
\email{lee.hw@northeastern.edu}
\orcid{0000-0001-6839-6257}
\affiliation{%
  \institution{Northeastern University}
  \city{Boston}
  \state{Massachusetts}
  \country{USA}
}

\author{Martina Di Paola}
\email{martinadipaola@kaist.ac.kr}
\orcid{0009-0005-8027-9834}
\affiliation{%
  \institution{KAIST}
  \city{Daejeon}
  \country{South Korea}
}

\author{Yoo Jin Hong}
\email{dbwk18@kaist.ac.kr}
\orcid{0009-0005-5961-4642}
\affiliation{%
  \institution{KAIST}
  \city{Daejeon}
  \country{South Korea}
}

\author{Quang-Huy Nguyen}
\email{22028077@vnu.edu.vn}
\orcid{0009-0006-7753-3843}
\affiliation{
\institution{VNU University of Engineering and Technology}
\city{Hanoi}
\country{Vietnam}
}

\author{Joseph Seering}
\email{seering@kaist.ac.kr}
\orcid{0000-0001-7606-4711}
\affiliation{%
  \institution{KAIST}
  \city{Daejeon}
  \country{South Korea}
}

\renewcommand{\shortauthors}{Lee et al.}

\begin{abstract}
LLM-based multi-agent simulations are a rapidly growing field of research, but current simulations often lack clear modes for interaction and analysis, limiting the "what if" scenarios researchers are able to investigate. In this demo, we define three core operations for interacting with multi-agent simulations: \textit{inject}, \textit{fork}, and \textit{compare}. \textit{Inject} allows researchers to introduce external events at any point during simulation execution. \textit{Fork} creates independent timeline branches from any timestamp, preserving complete state while allowing divergent exploration. \textit{Compare} facilitates parallel observation of multiple branches, revealing how different interventions lead to distinct emergent behaviors. Together, these operations establish a vocabulary that transforms linear simulation workflows into interactive, explorable spaces. We demonstrate this vocabulary through a commodity market simulation with fourteen AI agents, where researchers can inject contrasting events and observe divergent outcomes across parallel timelines. By defining these fundamental operations, we provide a starting point for systematic causal investigation in LLM-based agent simulations, moving beyond passive observation toward active experimentation.
\end{abstract}



\begin{teaserfigure}
  \includegraphics[width=\textwidth]{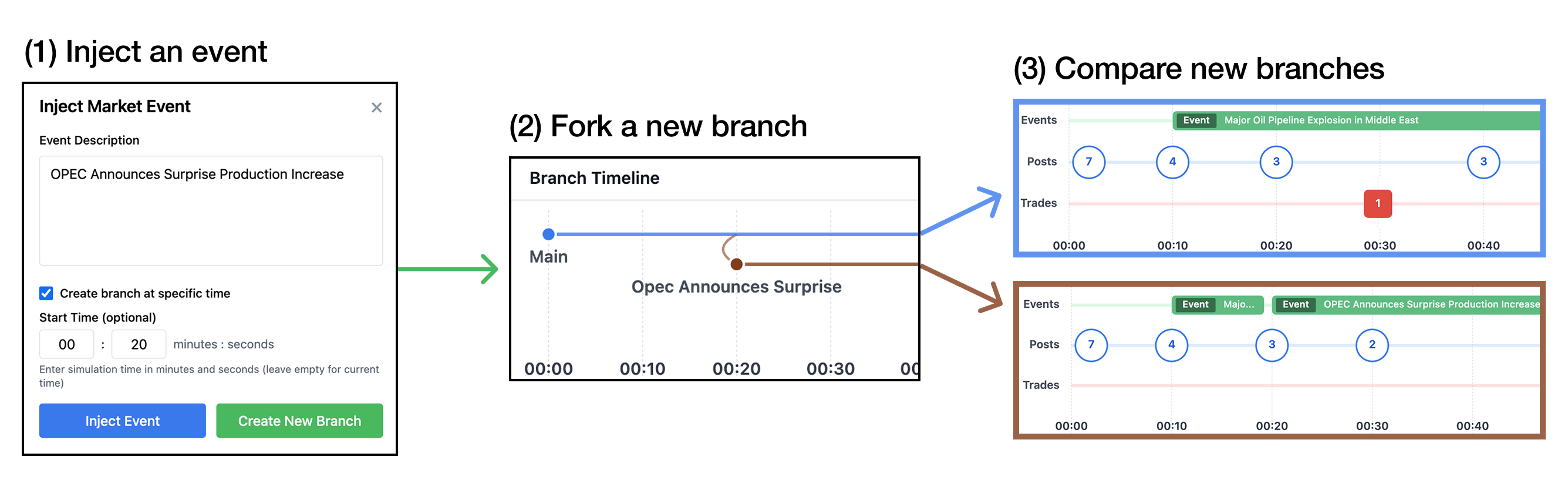}
    \caption{Workflow of our interactive space: (1) a popup component to \textit{inject} an event into the timeline or create a new branch at a selected time; (2) a Branch Timeline to \textit{fork} and visualize independent timeline branches; (3) an A/B Testing Dashboard to \textit{compare} and run two branches simultaneously.}
  \Description{Three main components of the interactive space are shown. First, a popup interface for injecting events or creating new branches at specific times. Second, a Branch Timeline that displays and organizes multiple timeline branches. Third, an A/B Testing Dashboard for comparing and running two timeline branches side by side.}
  \label{fig:teaser}
\end{teaserfigure}

\received{11 July 2025}

\maketitle

\section{Introduction}
The recent development of LLM-based multi-agent simulations has opened new opportunities for investigating complex social phenomena, with recent work demonstrating their promise as a research method~\cite{hu2025simulating, mou2024individual}. Researchers have successfully developed simulations across multiple domains, ranging from simulating large scale social media usage~\cite{zhang2025socioverse, rossetti2024social, touzel2024simulation}, to modeling rumor propagation~\cite{hu2025simulating} and understanding societal manipulation~\cite{anthis2025llm, mou2024individual, lin2025simspark}. These systems have shown promise in capturing emergent properties of social systems, demonstrating the potential for simulations to be used in deeper investigations of human behvaior at scale~\cite{anthis2025llm, zhang2025socioverse, park2022social, park2023generative}.

However, current simulation platforms remain fundamentally limited to single-path execution with limited interactability. Researchers must run complete simulations from start to finish for each hypothesis, making exploration time-consuming and preventing systematic comparison across scenarios~\cite{lin2025simspark}. Most critically, current platforms lack the ability to branch from past moments, preventing researchers from asking fundamental ``what if'' questions about how different interventions at critical points would have changed outcomes.

To address these limitations, we define an interaction vocabulary for exploring multi-agent simulations: \textit{inject}, \textit{fork}, and \textit{compare}. Our system demonstrates a set of novel interactions for researchers to use in navigating to any moment in a simulation's execution, branching into alternative timelines, and observing divergent outcomes in parallel. Unlike traditional workflows that require sequential runs with different parameters, our approach treats each moment as a potential branching point for systematic exploration. By establishing these operations, we allow researchers to isolate the impact of specific events and trace how different interventions lead to emergent behaviors, enabling real time hypothesis testing.

In our demo, we present a toy commodity market simulation with fourteen AI agents in a trading community with distinct strategies and market portfolio. Our demo presents the following scenario: starting from a stable market state, we inject contrasting events~---~for example, ``Major Oil Pipeline Explosion in Middle East'' in one branch and ``OPEC Announces Surprise Production Increase'' in another. The two scenarios are visualized side by side in real time, allowing researchers to watch how opposing events lead to divergent agent behaviors. This demonstrates how inject-fork-compare transforms simulation from passive observation into active investigation.

\section{Implementation}
Our implementation demonstrates the inject-fork-compare interaction vocabulary through a commodity market simulation with fourteen AI agents. Each operation is designed to enable researchers to systematically explore divergent outcomes from critical decision points.

Our demonstration features a dual-channel simulation where GPT-4o-mini powered agents operate simultaneously in social and market domains. In this environment, fourteen AI agents form a trading community, each with distinct investment strategies and market portfolios. The agents interact through two interconnected channels: they trade commodities in the market based on assigned strategies while simultaneously engaging in a social feed through posts and comments about currently active event(s). Though we use a commodity market for demonstration purposes, this dual-channel design represents a general pattern applicable to any domain where agents' decisions are influenced by environmental factors.\footnote{Note also that in this demo we focus on developing a set of interaction modalities rather than on developing a fully realistic market simulation; a toy simulation is sufficient for demonstrating these interactions here.}

\subsection{Inject: Event-Driven Hypothesis Testing}
The \textit{inject} operation allows researchers to introduce external events at any point during simulation execution, including retroactively at past timestamps. While SimSpark \cite{lin2025simspark} enables adding public events and monitoring simulations in real time, it operates within a forward-only execution model where events are added during ongoing simulations. Our implementation differs by combining event injection with temporal navigation. Researchers can navigate to any past moment and inject new events from that point. This retroactive injection capability transforms event injection from a linear configuration tool into an exploratory mechanism for testing alternative histories.

\subsection{Fork: Branching from Any Moment}
The \textit{fork} operation transforms a linear simulation into a branching exploration space. Similar to version control systems like Git\footnote{Git is a distributed version control system widely used in software development that allows developers to create branches from any point in a project's history, explore different features or fixes independently.} where developers can branch code to explore different features, our system allows researchers to branch simulations to explore different scenarios.

\begin{figure}[h]
\includegraphics[width=0.45\textwidth]{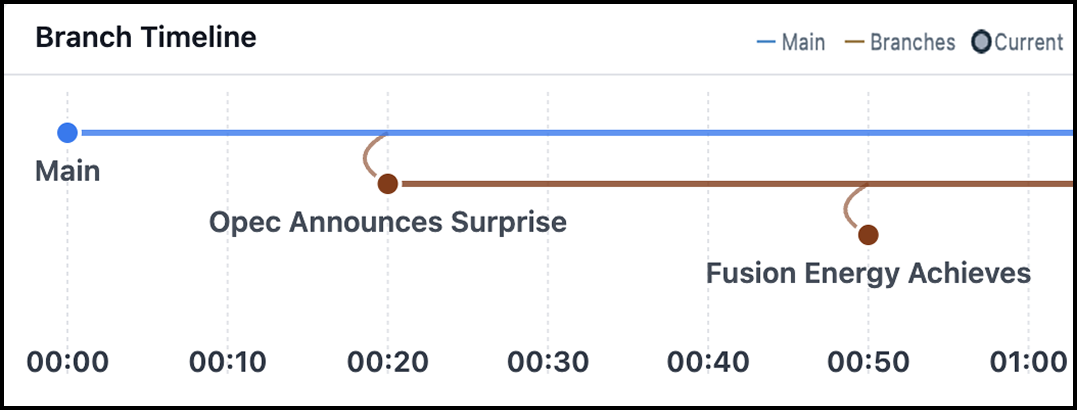}
\centering
\Description{The Branch Timeline interface displays multiple timeline branches. Users can fork a new branch from any point in the timeline, visualize each branch independently, and switch between them to explore different event sequences.}
\caption{A Branch Timeline to \textit{fork} and visualize independent timeline branches, and to switch between them for exploring alternative event sequences.}
\end{figure}

When a user initiates a fork, the system performs a complete state capture at that exact timestamp. This includes all market positions, social posts, and pending transactions. Everything in the captured state is then cloned to create an independent branch that inherits the complete history up to the fork point. From this moment forward, the branches evolve independently. Injected events lead to divergent agent behaviors and market outcomes. This mechanism ensures that differences between branches stem solely from the diverging events.

\subsection{Compare: Parallel Timeline Analysis}
The \textit{compare} operation transforms hypothesis testing into interactive exploration by enabling side by side observation of divergent simulation outcomes.

\begin{figure*}[t]
\centering
\includegraphics[width=0.85\textwidth]{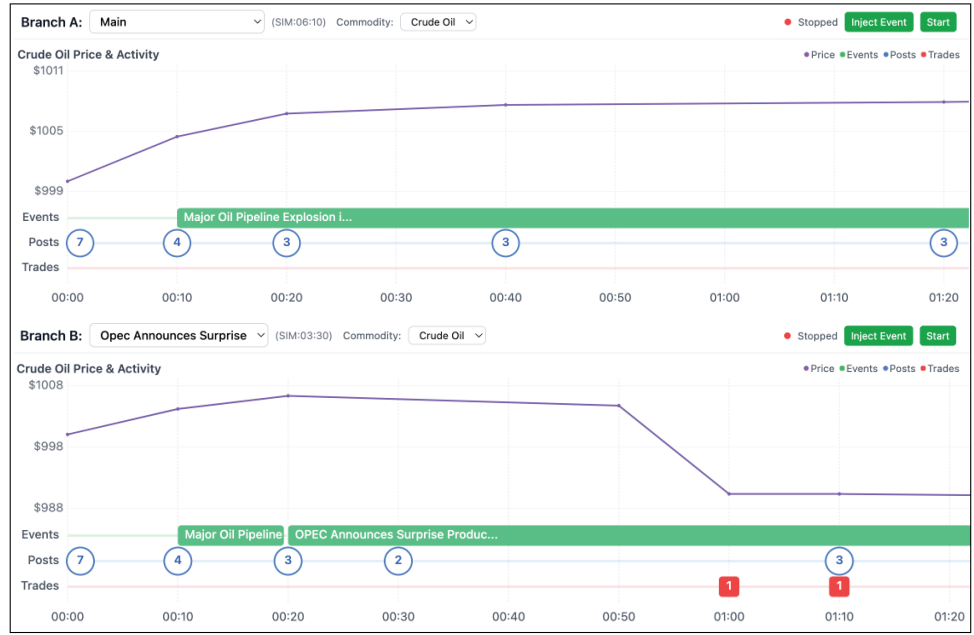}
\Description{The Parallel Control Interface displays two timelines next to each other. Users can switch between different branches and commodities within each timeline to compare them side by side.}
\caption{Parallel Control Interface displaying two timelines side by side, allowing users to switch between branches and commodities for flexible comparison.}
\label{fig:parallel}
\end{figure*}

\subsubsection{Activity Timeline}

Each simulation branch is visualized as an interactive timeline that overlays three key elements: injected events (green boxes), social activity (blue circles showing post counts), and market transactions (red markers showing transaction counts). Researchers can select different commodities to view their respective price movements. The green event boxes display which events were actively influencing agents and price at each timestamp, providing clear context for behavioral changes.

Interactive exploration reveals potential causal mechanisms behind agent decisions. Hovering over blue circles displays the titles of posts created at that moment, showing the social discourse triggered by events. Hovering over red markers exposes the reasoning behind each trade, i.e., why sellers decided to sell and buyers chose to buy. This detailed view allows researchers to trace how injected events cascade through social interactions into market actions.

\subsubsection{Parallel Control Interface}

The comparison interface places two timelines side by side, allowing researchers to observe how identical starting conditions diverge based on different injected events. Each timeline maintains independent controls for event injection and simulation management. Researchers can run both branches simultaneously to watch divergence unfold in real-time, or control them independently~---~pausing one branch to analyze critical moments while the other branch continues running.

\section{Future Work and Conclusion}
While our current implementation demonstrates inject-fork-compare through event injection, future work will expand the vocabulary of injectable elements. Beyond external events, the system could support injection of synthetic social content, allowing researchers to insert specific posts or comments to understand how individual-level changes propagate through complex systems. This interaction paradigm extends naturally to diverse simulation domains: urban planners could compare infrastructure interventions across parallel city evolutions, and epidemiologists could test containment strategies across branching disease spread models. While the ultimate success of these interactions will depend on the fidelity of each individual simulation, we argue that it is important to develop interaction modalities in parallel with research improving the accuracy of simulations at scale.

The inject-fork-compare vocabulary transforms how researchers interact with multi-agent simulations. Rather than running sequential experiments to answer "what if" questions, these interactions enable real-time exploration of alternative histories through branching. By allowing researchers to isolate specific interventions and observe their cascading effects across parallel timelines, we establish simulation as an active tool for causal investigation. As LLM-based simulations become increasingly central to understanding complex social phenomena, this ability to systematically navigate and manipulate temporal branches will prove essential for uncovering what happened and could have happened under different conditions.

\bibliographystyle{ACM-Reference-Format}
\bibliography{base}
\end{document}